%% file: apssamp.tex
\documentclass[aps,pra,twocolumn]{revtex4-2}

\usepackage{graphicx}% Include figure files
\usepackage{dcolumn}% Align table columns on decimal point
\usepackage{bm}% bold math
\usepackage{lineno}
\usepackage{amsmath}
\usepackage{float} 
\usepackage{placeins}
\usepackage{xcolor}
\begin{document}

\preprint{APS/123-QED}

\title{\textbf{Multiobjective Optimization for Robust Holonomic Quantum Gates} 
}% 

\author{Min-Hua Zhang$^{1}$}
\author{Jing Qian$^{1,2,3}$} 
 \email{Corresponding author: jqian1982@gmail.com}

\affiliation{$^{1}$State Key Laboratory of Precision Spectroscopy, Department of Physics, School of Physics and Electronic Science, East China Normal University, Shanghai, 200062, China
}
\affiliation{$^{2}$Chongqing Institute of East China Normal University, Chongqing, 401120, China}
\affiliation{$^{3}$Collaborative Innovation Center of Extreme Optics, Shanxi University, Taiyuan, 030006, China}

\date{\today}% It is always \today, today,

\begin{abstract}
The practical implementation of high-fidelity quantum gates faces significant challenges in simultaneously mitigating multiple operational errors arising from distinct physical mechanisms. These errors often span orders of magnitude in severity, and their respective suppression strategies may inherently conflict. In this work, we develop a universal multiobjective optimization framework for quantum gate design by integrating Pareto optimal solutions with an entropy-weight method. Using Rydberg-based nonadiabatic holonomic quantum gates (affected by amplitude errors, detuning errors, and Rydberg decoherence) as a testbed, we theoretically demonstrate the superiority of our algorithm. The optimized gates exhibit enhanced fidelity and robustness compared to those derived from one-objective optimization strategies. Furthermore, this framework is readily adaptable to other quantum gate protocols and provides a robust foundation for advancing fault-tolerant quantum computing.
\end{abstract}

\maketitle

%\tableofcontents

\section{Introduction}

Development of robust quantum gates is of great importance for scalable quantum computing with neutral atoms \textcolor{black}{\cite{robust_gate_Cong2022,robust_gate_Mitra2020,robust_gate_Mohan2023,robust_gate_Fromonteil2023,review_Shao2024}}, while inevitable errors that accumulate in practical quantum operations will fundamentally degrade the gate fidelity \textcolor{black}{\cite{review_Shi2022,gate_error_Pagano2022,errors_Guo2020}}. These errors stem from versatile noise sources including the technical imperfections and the decoherence inherent to the atoms \textcolor{black}{\cite{error_stem_Zhang2012}}. Recent efforts have shown that the execution of modulated pulses based on optimal control can realize high-fidelity entangling gates \textcolor{black}{\cite{Modulated_Pulses_Fidelity_Levine2019,Modulated_Pulses_Fidelity_Fu2022,Nature_Evered2023}}, and meanwhile make the system more insensitive to the imperfection of operations \textcolor{black}{\cite{Modulated_Pulses_Robust_Daems2013,Modulated_Pulses_Robust_Goerz2014,Modulated_Pulses_Robust_Poggi2024,Modulated_Pulses_Robust_Zhang2024,Modulated_Pulses_Robust_Goerz2014_PRR}}.

However, conventional optimization schemes focuses on one objective, such as solely pursuing a maximal gate fidelity in the absence of any error \textcolor{black}{\cite{Modulated_Pulses_Fidelity_Theis2016}} or taking account of the robustness to one specific error type \textcolor{black}{\cite{One_objective_Robust_Fauseweh2012,One_objective_Robust_Hou2024,One_objective_Robust_Xiao2024}}. In reality, these existing errors are usually conflicting objectives so optimizing one type of errors should require the cost of the others \textcolor{black}{\cite{decline_other_objective_Liang2023}}. A recent study proposes a novel pulse that is robust against both the amplitude and the detuning deviations by simply adding two cost functions into one objective in optimization \textcolor{black}{\cite{Two_Cost_Function_into_One_PRX_Jandura2023}}. This method is applicable only when the two errors have comparable magnitudes. Whereas if they span orders of magnitudes in severity, the one-objective optimization strategy will be biased towards the dominant objective with larger magnitude resulting in a failure in obtaining the balanced optimization. Consequently, developing multiobjective optimization algorithm \textcolor{black}{\cite{Multi_objective_optimization_Sharma2022,Multi_objective_optimization_Gollub2009,Multi_objective_optimization_Chatterjee2025,Multi_objective_optimization_chen2025modl}} for achieving robust gates against versatile coexisting errors, remains a critical yet unresolved challenge for neutral-atom quantum computing.

In this work, we present a novel multiobjective optimization framework for the implementation of nonadiabatic holonomic quantum computation (NHQC) \cite{NHQC_Review_Zhang2023}, \textcolor{black}{a class of geometric quantum gates that features rapid non-Abelian evolution and inherent resilience due to the removal of dynamical phase contributing a pure geometric nature \cite{NHQC_revised_geometric_condition_Sjqvist2012,NHQC_revised_Xu2012}, which has been further exploited with diverse physical platforms in recent years  {\cite{NHQC_AbdumalikovJr2013,NHQC_Feng2013,NHQC_Zhou2017,NHQC_Zhao2020,NHQC_Jin2024}}. Despite these advantages, the geometric gate remains susceptible to the decoherence processes and control imperfections inherent in practical quantum systems, particularly amplitude fluctuations and detuning errors.} Such perturbations can induce noncyclic evolution paths and accumulate nonvanishing dynamical phases, thereby degrading gate fidelity and operational robustness \textcolor{black}{\cite{Control_error_PhysRevA.95.012334,Control_error_Thomas2011,Control_error_Zheng2016}}. Prior approaches to enhance NHQC performance have predominantly relied on one-objective optimization strategies, by 
precisely modifying the path parameters in the time domain \textcolor{black}{\cite{Precisely_Modify_the_Path_Parameters_Liang2024}} or using cleverly-designed composite dynamical decoupling pulses \textcolor{black}{\cite{Composite_Dynamical_Decoupling_pulses_Liang2022}}, which undoubtedly 
add to the difficulty in practice.

To address the challenge we apply multiobjective optimization by designing a set of cost functions 
for independent optimization goals which  simultaneously minimize three major imperfections in NHQC: amplitude fluctuation, detuning deviation and Rydberg decoherence. Our algorithm can 
effectively coordinate the conflicting multiple objectives and ultimately generate a Pareto-optimal solution set \cite{Pareto_front_Haxhiraj2025}. Furthermore, to identify the most practical optimal solution even with significant magnitude differences among the cost functions, we employ the entropy-weight method (EWM) \textcolor{black}{\cite{EWM_Kumar2021}} which determines the optimal weight allocation. Based on the robust pulses we demonstrate superior fidelity and robustness in constructing single- and two-qubit geometric gates as compared to one-objective optimized gates. By providing a systematic approach to multi-parameter trade-off analysis, our work is generalizable to other protocols strongly advancing fault-tolerant quantum gate engineering.

% The construction of decoherence-free subspaces have been proposed specifically designed for decoherence suppression.\textcolor{blue}{[add ref]}. These methods share a fundamental limitation that only optimize individual error sources while failing to simultaneously handle the three primary error types - decoherence, detuning, and amplitude errors leading to significantly degraded gate performance in realistic complex environments compared to theoretical predictions.

   % Furthermore, it generally demonstrate inferior robustness against control errors when compared to conventional dynamical quantum gates\textcolor{blue}{\cite{Control_error_Zheng2016,Control_error_Thomas2011,Control_error_Ramberg2019,Control_error_PhysRevA.95.012334}}.

\section{Theoretical strategy}

\begin{figure}
\includegraphics[width=3.4in]{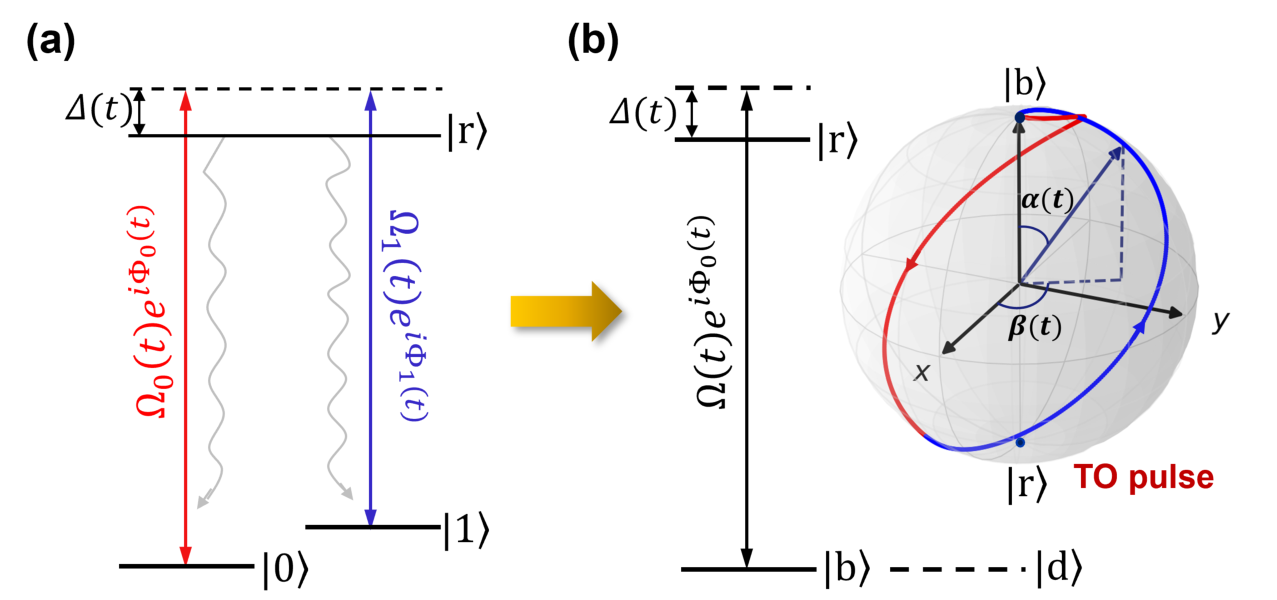}
\caption{\label{modone} Basic idea of single-qubit nonadiabatic holonomic quantum gates. (a) Energy levels for a three-level atom, where the qubits are encoded in ground hyperfine states ${\vert 0 \rangle},\vert 1\rangle$, and the Rydberg state $\vert r \rangle$ acts as an auxiliary level that spontaneously decays to the ground states. The ground-Rydberg transitions between $|r\rangle$ and $|0\rangle$ (or $|1\rangle$) are realized by off-resonant two-photon laser fields with a common detuning {$\Delta(t)$}.
(b) Effective two-level system in the $\{|b\rangle,|r\rangle\}$ subspace where the dark state $|d\rangle$ is decoupled. Inset: Illustration of the second dark path $|\mu_2(t)\rangle$ evolution with a close loop
on the Bloch sphere of $\{|b\rangle,|r\rangle\}$ subspace, representing the realization of single-qubit gate operation based on the TO pulse.}
\end{figure}

To make the statement clear, we first review the realization of single-qubit gates in NHQC \textcolor{black}{\cite{NHQC_revised_geometric_condition_Sjqvist2012}}. The typical scheme (Fig.\ref{modone}a) considered contains a three-level atom with two ground states $\vert 0 \rangle$, $\vert 1\rangle$ and one Rydberg state $\vert r\rangle$. The coherent excitation between $\vert j \rangle$ and $\vert r\rangle$ is enabled by a two-photon laser pulse with time-dependent Rabi frequency $\Omega_{j} (t) e^{i\Phi_{j}(t)}$ ($j=0,1$) and a same detuning ${\Delta(t)}$. In the interaction frame the Hamiltonian of the system is
\begin{eqnarray}\label{eq1}
{ {H}(t)}=&&
\frac{1}{2}\left[\Omega_{0} (t) e^{-i\Phi_{0}(t)}{\vert 0 \rangle}{\langle r \vert}  
\nonumber + \Omega_{1} (t)e^{-i\Phi_{1}(t)}{\vert 1 \rangle}{\langle r \vert} \nonumber +\text{H.c.}\right]
\\&& -\Delta(t){\vert r\rangle} {\langle r\vert}.
\end{eqnarray}

To construct an arbitrary single-qubit holonomic gate, the pulse shapes should be parametrized as \textcolor{black}{\cite{Hamiltonian_Ai2022}}
\begin{equation}
 \Omega_{0} (t)=\Omega(t)\sin{\frac{\theta}{2}},\Omega_{1} (t)=\Omega(t)\cos{\frac{\theta}{2}}  
\end{equation}
and $\Phi=\Phi_{0}(t)-\Phi_{1}(t)$ forms a time-independent phase difference. Note that $|0\rangle,|1\rangle$ are two qubit states and $|r\rangle$ acts as an auxiliary state. Therefore, by defining 
\begin{equation}
 \vert b \rangle=\sin{\frac{\theta}{2}}\vert 0 \rangle+\cos{\frac{\theta}{2}}e^{-i\Phi}\vert 1 \rangle, \vert d\rangle=\cos{\frac{\theta}{2}}e^{i\Phi}\vert 0 \rangle-\sin{\frac{\theta}{2}}\vert 1 \rangle  \nonumber 
\end{equation}
the effective Hamiltonian can be recast as
\begin{equation}
  {H}_{b}(t)=\frac{\Omega(t)}{2}e^{-i\Phi_{0}(t)}{\vert b \rangle}{\langle r \vert} +\text{H.c.} -\Delta(t) {\vert r\rangle} {\langle r\vert}
\end{equation}
in the $\{|b\rangle,|r\rangle\}$ subspace (Fig.\ref{modone}b), while the dark state $|d\rangle$ with a zero eigenenergy is decoupled, i.e. $H(t)|d\rangle=0$. In this way, the dark state $|d\rangle$ remains unchanged for constants $\theta,\Phi$ during the quantum dynamics leading to a zero dynamical phase, undoubtedly unsuitable for a geometric phase gate.

We proceed to find out another dark path for implementing the gate. In the $\{|b\rangle,|r\rangle\}$ subspace one choice of auxiliary vectors is
\begin{eqnarray}
 {\vert \mu_{2}(t)\rangle} &=& {\cos{\frac{\alpha(t)}{2}}\vert b\rangle + \sin{\frac{\alpha(t)}{2}}}e^{i\beta(t)}{\vert r \rangle} \\
 {\vert \mu_{3}(t)\rangle} &=& {-\sin{\frac{\alpha(t)}{2}}}e^{-i\beta(t)}\vert b\rangle + \cos{\frac{\alpha(t)}{2}}{\vert r \rangle}
\end{eqnarray}
and $|\mu_1(t)\rangle=|d\rangle$. Differing from the first dark state $|\mu_1(t)\rangle$ we require the second dark path could induce a cyclic evolution towards an expected geometric phase. Here, we select ${\vert \mu_{2}(t)\rangle}$ to be the other dark path and exclude ${\vert \mu_{3}(t)\rangle}$ because it is initially unoccupied for the choice of $\alpha(0)=0$ (see Eq.\ref{pra}). Now we should ensure the removal of the dynamical phase in $|\mu_2(t)\rangle$ at any time during the evolution, by letting 
\begin{equation}
    {\langle \mu_{2}(t)\vert} H_{b}(t){\vert \mu_{2}(t)\rangle}=0 \label{par}
\end{equation}
which is named as the parallel transport condition. As a consequence, the evolution is purely geometric with vanishing dynamical contribution.
Besides, the other cyclic evolution condition must be satisfied at the same time, i.e. ${\vert \mu_{2}\small{(\tau)}\rangle}{ \langle\mu_{2}\small{(\tau)}\vert}={\vert \mu_{2}(0)\rangle} {\langle\mu_{2}(0)\vert} $ with $\tau$ the gate duration, arising the boundary condition $\alpha(\tau)=\alpha(0)=0$ for $\alpha$ function, which ensures that the dark path $|\mu_2(t)\rangle$ returns to $|b\rangle$ at $t=\tau$. Therefore, we can finally obtain a pure geometric phase for the targeted nonadiabatic holonomic quantum gates, which is
\begin{equation}
    \gamma(\tau) =i\int_{0}^{\tau}{\langle \mu_{2}(t)\vert\dot{\mu}_{2}(t)\rangle} dt =\frac{1}{2}\int_{0}^{\tau}\dot{\beta}(t)\left[ 1-\cos{\alpha(t)} \right]dt \label{phase}.
\end{equation}

Consequently, the whole holonomic matrix of the
geometric evolution can be given by
\begin{equation}\label{U}
   U=\vert d\rangle\langle d\vert+e^{i\gamma(\tau)}\vert b\rangle\langle b \vert
\end{equation}
in the effective $\{\vert d \rangle, \vert b \rangle\}$ subspace, which can also be rewritten in the initial computational basis of $\{\vert 0 \rangle, \vert 1 \rangle\}$ as 
\begin{eqnarray}\label{Ut}
  \small
  &&\small
  U\small{(\theta,\Phi,\gamma)} = \left[
  \begin{array}{ccc}
   \cos^{2}{\frac{\theta}{2}}+e^{i\gamma}\sin^{2}{\frac{\theta}{2}} 
  &  \frac{1}{2}\sin{\theta}e^{i\Phi}\left(e^{i\gamma}-1 \right)
  \\
   \frac{1}{2}\sin{\theta}e^{-i\Phi}\left(e^{i\gamma}-1 \right)
  &  \sin^{2}{\frac{\theta}{2}}+e^{i\gamma}\cos^{2}{\frac{\theta}{2}}
  \end{array}
  \right].\nonumber\\&&
\end{eqnarray}
Using the unitary matrix $U(\theta,\Phi,\gamma)$ in Eq.(\ref{Ut}) we can construct arbitrary single-qubit phase gate. For example, $X=U(\frac{\pi}{2},\pi,\pi)$, $H=U(\frac{\pi}{4},\pi,\pi)$, $T=U(0,\pi,\frac{\pi}{4})$ and $S=U(0,\pi,\frac{\pi}{2})$.
In this work we focus on the $X$ gate which is particularly useful for a two-qubit CNOT gate while incorporating with a control qubit (see Sec. VB) \textcolor{black}{\cite{CNOT_gate_Isenhower2010}}. 

Furthermore, according to the von Neumann equation $i\dot{\rho}_{j}(t) = [H_b(t),\rho_{j}(t)] $ with $\rho_{j}(t) = {\vert \mu_{j}(t)\rangle}{\langle \mu_{j}(t)\vert} $ $(j=1,2)$ \textcolor{black}{\cite{von_Neumann_equation_Liu2019}} alongside with the parallel transport condition in Eq.(\ref{par}), the laser and detuning parameters can be reversely determined by
\begin{eqnarray}\label{RePa}
     &&\Omega(t) =\frac{\dot{\alpha}(t)}{\sin{\left[\Phi_{0}(t)-\beta(t)\right]}}\nonumber, \\&&
    \Phi_{0}(t)= \beta(t)-\arctan{\left[ \frac{\dot{\alpha}(t)\cot{\alpha(t)}}{\Delta(t)+\dot{\beta}(t)}\right]},\\&&
    \Delta(t)=-\dot{\beta}(t)\left[1+\cos{\alpha(t)}\right].\nonumber
\end{eqnarray}
It is explicit that with an appropriate set of variables $\alpha(t)$ and $\beta(t)$ the effective Hamiltonian $H_b(t)$ can be inversely engineered towards a desired evolution. For example, one path of evolution state $\vert \mu_{2}(t) \rangle$ given by the TO pulse, is visualized on the Bloch sphere in Fig.\ref{modone}b ensuing the cyclic evolution,  where $\alpha(t)$ and $\beta(t)$ serves as the time-dependent polar and azimuth angles, respectively. Note that, to realize the $X$ gate, the original pulse parameters are $\Omega_0(t)=\Omega_1(t)=\frac{\sqrt{2}}{2}\Omega(t)$ and $\Phi_1(t) = \Phi_0(t)-\pi$.

\section{One-objective Optimization}

\subsection{Non-robust pulses under typical optimization}

\begin{table*}[t] % 使用 table* 环境，表格放置在页面顶部
\caption{\label{tab:table1}Optimized parameters for different types of pulses which lead to the implementation of single-qubit $X$ gates. }
\begin{ruledtabular}
\renewcommand{\arraystretch}{1.2} % 调整行高为默认值的1.5倍
\begin{tabular}{ccccccc}
 & \multicolumn{3}{c}{Parameters for $\alpha(t)$} & \multicolumn{3}{c}{Parameters for $\beta(t)$} \\
 \cline{2-4} \cline{5-7}
 Pulse Type & $a_{\alpha}$ & $b_{\alpha}$ & $c_{\alpha}$
 & $a_{\beta}$ & $b_{\beta}$ & $c_{\beta}$
  \\ \hline
 TO & $2.4630$ & $1.0$ & $4.0$ & $17.2074$ & $0.2154$ & $1.0$  \\ 
 AR & $-3.3049$ & $4.0$ & $2.0$ & $-5.9809$ & $0.5187$ & $2.0$  \\ DR & $-3.4016$ & $1.0$ & $1.0$ & $3.3443$ & $0.5746$ & $5.0$ \\
 ADR &$-3.1678$ & $1.0$
 &$2.0$ & $29.3233$ & $0.9620$ & $2.0$ \\
 SR &$-2.6166$ & $1.0$ & $5.0$ & $-20.0805$ & $0.9556$ & $5.0$   \\
\end{tabular}
\end{ruledtabular}
\end{table*}

We proceed to evaluate the performance of single-qubit gates using the Lindblad master equation \textcolor{black}{\cite{Lindblad_master_equationBraaten2017}}
\begin{equation}\label{master}
\frac{d {\rho}}{dt}=-i[H(t), {\rho}]+{\mathcal{L}_r}[\rho]+{\mathcal{L}_z}[\rho]
\end{equation}
with $\rho$ the density matrix and $\mathcal{L}_{r,z}$ the Lindblad superoperators, representing the decoherence from spontaneous emission of $|r\rangle$ and a global dephasing due to laser phase noise, 
which take forms of
\begin{eqnarray}
    {{\mathcal{L}}_{r}[{\rho}]} & {=} & \sum_{j=0,1}\left[{L}_{j}{\rho}{L}_{j}^\dagger - \frac{1}{2}\textcolor{black}{\{{L}_{j}^\dagger {L}_{j}, {\rho}\}}\right] \nonumber\\
    {\mathcal{L}}_{z}[{\rho}] &=& {L}_{z}{\rho}\hat{L}_{z}^\dagger - \frac{1}{2}\{{L}_{z}^\dagger {L}_{z}, {\rho}\}
\end{eqnarray}
with ${{L}}_{j} = \sqrt{\Gamma_r/2}\ \vert j \rangle \langle r \vert$ and ${{L}_{z}}=\sqrt{\Gamma_{z}}(\vert 0 \rangle\langle 0 \vert+\vert1\rangle\langle1\vert-\vert r\rangle\langle r\vert)$. Here, $\Gamma_r$ and $\Gamma_z$ present the spontaneous decay and dephasing rates. \textcolor{black}{To quantify the gate performance, we adopt the fidelity function by taking account of the average effect of two input states $|0\rangle,|1\rangle$, which can faithfully measure the accuracy how the realized gate operation matches the ideal one} \cite{Fidelity_Nielsen2012}, 
\begin{equation}\label{eq7}
\mathcal{F}=\frac{1}{2}\text{Tr}\left[ \sqrt{\sqrt{U} \rho(\tau) \sqrt{U} } \right],
\end{equation}
where $U=U(\frac{\pi}{2},\pi,\pi)$ is the ideal holonomic transformation for $X$ gate and $\rho(\tau)$ is the realized density matrix at $t=\tau$ of evolution.

We then turn to implement the $X$ gate with a guess for two specific functions \textcolor{black}{\cite{Light_Liang2024}}
\begin{equation}
 \alpha (t)=a_{\alpha}\left[\sin{\left(\frac{b_{\alpha}\pi t}{\tau}\right)}\right]^{c_{\alpha}}, \beta (t)=\ a_{\beta} \left[\sin{\left(\frac{b_{\beta}\pi t}{\tau}\right)}\right]^{c_{\beta}}   \label{pra}
\end{equation}
to ensure the cyclic evolution condition. Due to a variety of tunable parameters $(a_{\alpha}, b_{\alpha}, c_{\alpha};a_{\beta}, b_{\beta}, c_{\beta})$ we perform a global optimization for the waveforms of $\alpha(t),\beta(t)$ targeting at minimizing the infidelity $1-\mathcal{F}$. To ease the optimization procedure we also note that $b_\alpha$ must take integer values limited by the boundary condition $\alpha(\tau)=\alpha(0)=0$, so set $b_\alpha$ to be adjustable among integers in $(0,5]$, the same for $c_\alpha,c_\beta$ which are exponential terms. To prevent the accumulated phase $\gamma$ from evaluating to be zero in Eq.(\ref{phase}), it is better to preserve the asymmetry of $\beta$ function with respect to $t=\tau/2$ so $b_\beta$ is tunable within the range of $(0,1)$. The remaining $a_\alpha,a_\beta$ coefficients are freely adjustable within a wide range of {$[-30,30]$} and {$\tau=1.0$ $\mu$s}.

To our knowledge, the performance of any quantum gate should be practically measured by the gate fidelity subject to external noises. However, holonomic gates cannot be significantly robust to these control errors that will change the designed evolution paths \textcolor{black}{\cite{change_designed_path_Liu2021}} or lead to a nonzero dynamical phase \textcolor{black}{\cite{Precisely_Modify_the_Path_Parameters_Liang2024}}. 
Here, we parametrize the operational imperfections that impact the gate in two ways. One is the laser amplitude fluctuations that arise an uncertain Rabi frequency of $(1+\epsilon)\Omega(t)$. The other is the laser frequency error that can result in a detuning deviation $(1+\eta)\Delta(t)$ to the two-photon transition. We also note that the intrinsic decoherence from the auxiliarily Rydberg excited state $|r\rangle$ can lead to gate infidelity, characterized by \textcolor{black}{$\kappa = \Gamma_z = 10\Gamma_r$, since the Rydberg dephasing predominantly induced by laser phase noise exceeds the spontaneous decay rate by approximately an order of magnitude \cite{decay_dephasing_Tamura2020}}. Thereby, the original Hamiltonian (\ref{eq1}) can be re-expressed in a more universal form
\begin{eqnarray}
H(t) &=&  \frac{(1+\epsilon)}{2}[ \Omega_0(t)e^{-i\Phi_0(t)}|0\rangle\langle r| + \Omega_1(t) e^{-i\Phi_1(t)}|1\rangle\langle r|  \nonumber\\
&+& \text{H.c.}] - (1+\eta)\Delta(t)|r\rangle\langle r|
\label{eqfull}
\end{eqnarray}
where $\epsilon$ and $\eta$ are small deviations for the external errors. Note that we assume same amplitude error $\epsilon$ for $\Omega_0(t)$ and $\Omega_1(t)$,
\textcolor{black}{because the realistic lasers driving the ground-Rydberg transitions differ by a small frequency of several GHz (corresponding to the hyperfine splitting energy of two ground states) which can be practically realized by a single laser with sideband modulation \cite{Same_amplitude_error_McDonnell2022}. Consequently a common detuning error $\eta$ is adopted representing the imperfection in laser frequency.}

\begin{figure}
\includegraphics[width=3.5in]{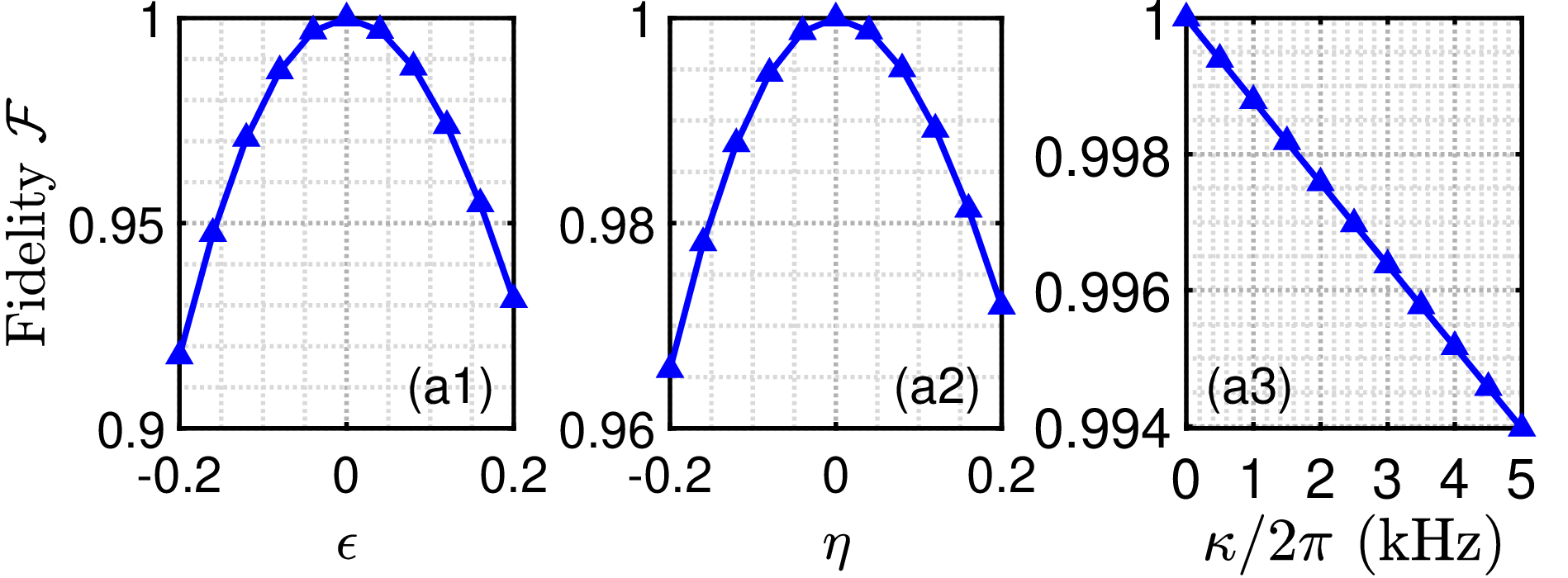}
\caption{\label{errs} Gate performance based on the typical one-objective optimization with respect to (a1) the $\epsilon$ error, (a2) the $\eta$ error and (a3) the $\kappa$ error.}
\end{figure}

To demonstrate the gate performance, we first utilize a typical one-objective numerical optimization with the genetic algorithm as in our previous work \textcolor{black}{\cite{previous_work_Li2024}}, merely targeting at minimizing the average infidelity in the absence of any error (refer as the typical-optimal or TO), in which the cost function is given by
\begin{equation}
    \mathcal{J}_{to} = 1- \mathcal{F}_{(\epsilon,\eta,\kappa)=0}
    \label{tocost}.
\end{equation}
Note that the TO pulse does not have any robustness because the cost function (\ref{tocost}) excludes the consideration for any error. \textcolor{black}{All optimized parameters $(a_\alpha,b_\alpha,c_\alpha;a_\beta,b_\beta,c_\beta)$ will be used to determine the functions of $\alpha(t)$ and $\beta(t)$, which are subsequently mapped to the complete set of physical control parameters ($\Omega(t)$, $\Phi_0(t)$, $\Delta(t)$) as required for the implementation of quantum gates.}

Alongside with the optimized parameters described in Table \ref{tab:table1}, the TO pulse under zero-noise environment can achieve a perfect infidelity of $1-\mathcal{F}_{(\epsilon,\eta,\kappa)=0}< 10^{{-14}}$ because of the parallel transport and boundary conditions, confirming the realization of $X$ gates. Remarkably, we have restricted the maximal laser amplitude to be $2\pi\times 10$ MHz for a practical two-photon coupling strength \textcolor{black}{\cite{two_photon_coupling_Cantu2020}} that leads to a relatively small $|a_\alpha|$ value due to $\Omega(t)\propto \dot{\alpha(t)}$.
Fig. \ref{errs} comparably show the $X$ gate fidelity with respect to different $\epsilon,\eta,\kappa$ errors. In general, such gates are very sensitive to the external control and intrinsic errors. The fidelity values, although achieving $\mathcal{F}\approx 1$ at $(\epsilon,\eta,\kappa)=0$, 
will suffer from a dramatic decrease as long as any error presents. Nevertheless, a comparison among (a1-a3) also presents that the geometric gates are more robust to the decoherence error ($\kappa$) than other environment noises ($\epsilon,\eta$).
Numerical results indicate that, for very large values of $|\epsilon|$ up to 0.2 the infidelity of the TO pulse has attained {${\sim 0.0824}$}. A comparable magnitude level of reduction in fidelity is also found for the laser frequency error which achieves {${\sim 0.0343}$} at $|\eta|=0.2$.
While, the gate fidelity reveals a linear decrease as the decoherence error $\kappa$ grows which arises the infidelity of \textcolor{black}{$0.0060$} when $\kappa/2\pi=5$ kHz, smaller than the $\epsilon,\eta$ errors by one order of magnitude. Notably, such a substantial difference in magnitudes of multiple errors will add to difficulty for one-objective optimization.
We check different imperfections to the infidelity simultaneously, and estimate the average fidelity as $\mathcal{F} \approx \textcolor{black}{0.8773}$ conservatively, for single-qubit $X$ gates in a practical environment with strong noises $(\epsilon,\eta,\kappa/2\pi)=(0.2,0.2,5.0 \text{ kHz})$.

%It follows that robustness optimization of quantum gates against errors must be taken into account. To achieve high-fidelity gates in a real environment with coexisting multiple types of noise, we next construct a super-robust NHQC using a powerful multi-objective optimization strategy to simultaneously enhance robustness against various types of errors. 

\subsection{Amplitude- and Detuning-robust pulses} 

In this subsection, we start by finding robust pulses that are more insensitive to the laser amplitude ($\epsilon$) or the detuning ($\eta$) deviations. In experiment, types of errors occur in the Hamiltonian which could make the realized gate deviate from the target one \textcolor{black}{\cite{errors_Levine2018}}. For example, the presence of $\epsilon$ or $\eta$ errors can lead to a significant infidelity $\sim 10^{-2}$ with the TO pulse, larger than the decoherence error ($\kappa$) by one order of magnitude. To achieve a robust pulse against the dominant amplitude or detuning deviations, we first modify the cost function by
\begin{equation}
   \mathcal{J}_{ar}=1-\frac{\mathcal{F}{(\epsilon_{1})} + \mathcal{F}{(\epsilon_{2})} + \cdots + \mathcal{F}{(\epsilon_{N})}}{N} \label{ar}
   \end{equation}
for an amplitude error robust (AR) pulse and 
\begin{equation}
   \mathcal{J}_{dr}=1-\frac{\mathcal{F}{(\eta_{1})} + \mathcal{F}{(\eta_{2})} + \cdots + \mathcal{F}{(\eta_{N})}}{N} \label{dr}
\end{equation}
for a detuning robust (DR) pulse, and individually minimize them using the genetic algorithm. \textcolor{black}{By performing the optimization, e.g. for a AR pulse, we assume a wide error range of $\epsilon_{i} \in [-\epsilon_{0}, \epsilon_{0}]$ with $i=1, 2,  \ldots, N$ ($\epsilon_{i}$ is uniformly distributed with 
$\epsilon_{0}=0.2$ the maximal deviation degree and $N=20$). With the prior knowledge of error strength, in each optimization epoch we could solve every fidelity value $\mathcal{F}(\epsilon_{i})$ under different error rate $\epsilon_i$, so as to evaluate the cost function $\mathcal{J}_{ar}$. The maximization of the average gate fidelity over every fidelity value $\mathcal{F}(\epsilon_i)$ with an error rate $\epsilon_i$ ($\eta=\kappa=0$) will give rise to the AR pulse.} Note that the DR pulse is obtained by using the same way.

\begin{figure}
\includegraphics[width=3.5in]{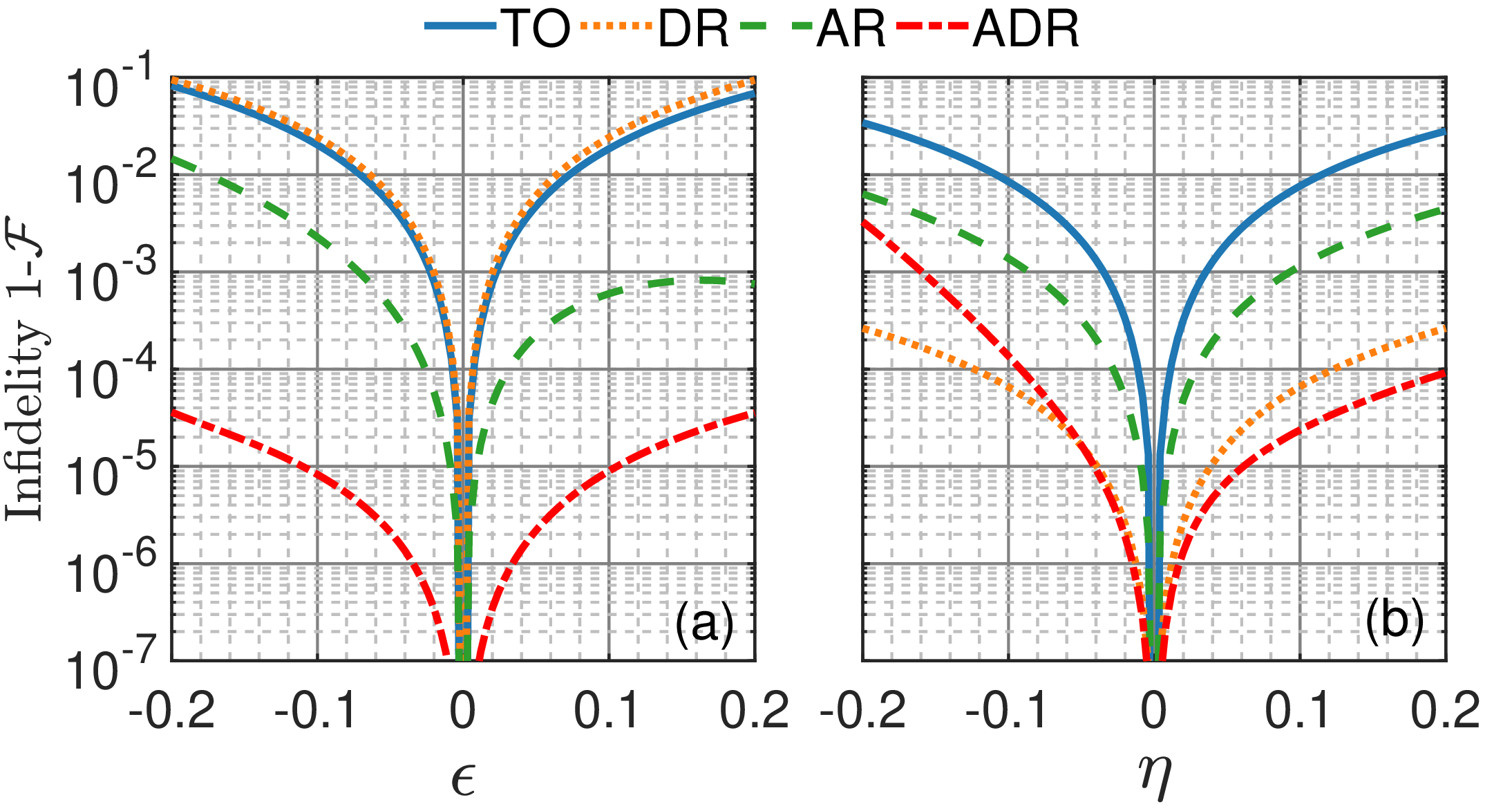}
\caption{\label{different_pulse} Dependence of the gate infidelity $1-\mathcal{F}$ of different pulses in the absence of decoherence $\kappa=0$ on (a) the amplitude error $\epsilon$ with $\eta=0$ and (b) the detuning error $\eta$ with $\epsilon=0$.}
\end{figure}

To explore the robustness of the AR and DR pulses, we calculate the infidelity $1-\mathcal{F}$ in the absence of $\kappa$ error in Fig.\ref{different_pulse}(a) and(b),
as a function of the $\epsilon$ and $\eta$ errors respectively. We observe that the AR pulse can exhibit a significant improvement in infidelity against the deviation of amplitude errors, outperforming the TO and DR pulses by orders of magnitude. In particular, when $\epsilon$ is positive that leads to a larger laser amplitude the gate is more robust to the amplitude errors. While the DR pulse does not have any robustness to $\epsilon$, behaving similarly as the TO pulse. However, the DR pulse can achieve an infidelity of {$1-\mathcal{F} \lesssim 3\times 10^{-4}$} even for $|\eta|$ up to 0.2, significantly improving on the TO and AR pulses. Restricted by the one-objective optimization algorithm, although
an explicitly enhanced robustness of the AR and DR pulses (as compared to TO) occurs in their respective targets, we have to admit that this superiority arises at the cost of sacrificing robustness to the other type of error as observed in Ref.\textcolor{black}{\cite{Two_Cost_Function_into_One_PRX_Jandura2023}}. It is more promising to seek for optimized parameters that can strengthen the gate robustness against both types of errors $\epsilon,\eta$.

\section{Multiobjective optimization}

\subsection{Design of the ADR pulse using EWM}

\begin{figure}
\includegraphics[width=3.5in]{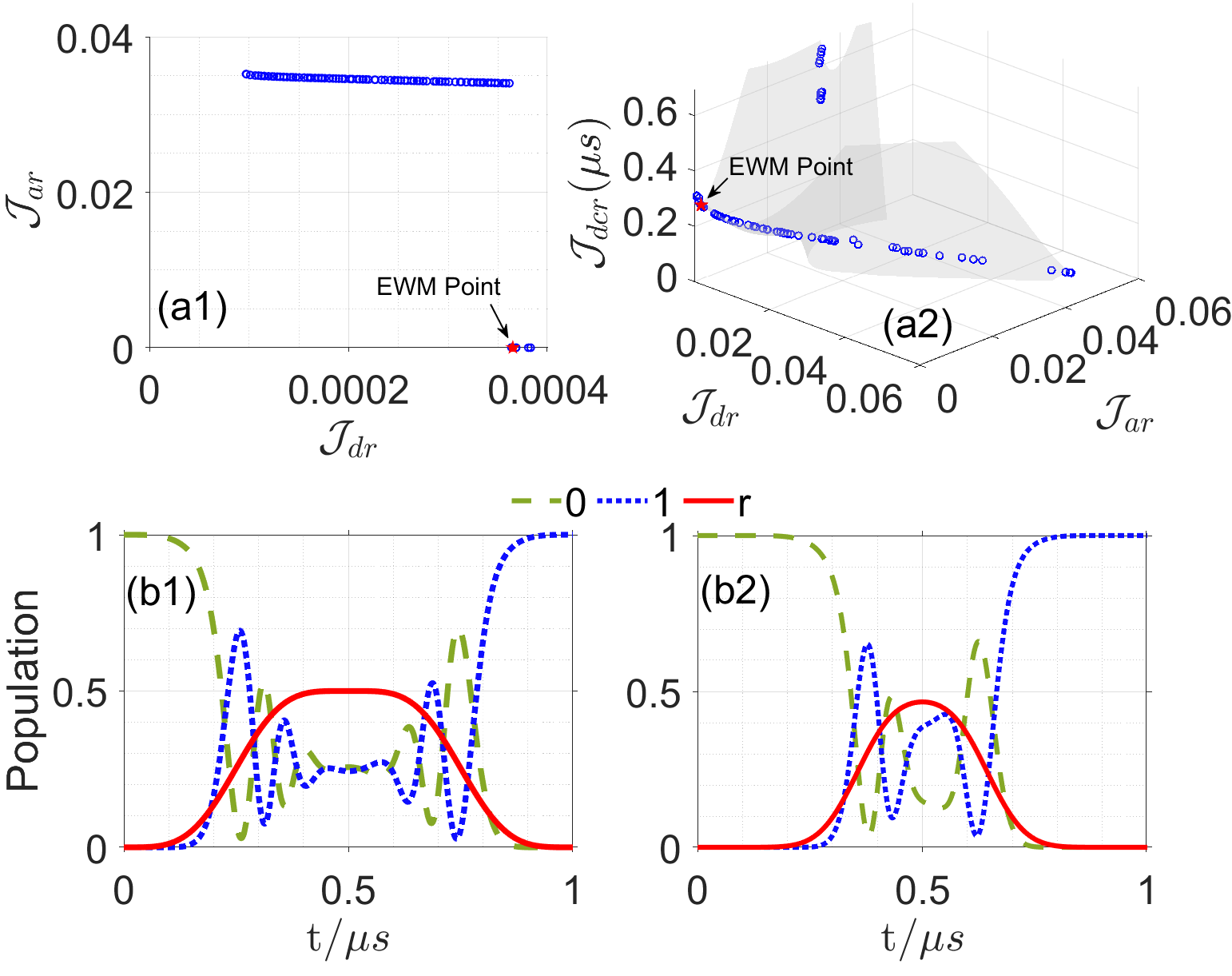}
\caption{\label{ParetoFront} (a) PF formed by sufficient Pareto points ($n=100$) for (a1) two objectives ($\mathcal{J}_{dr},\mathcal{J}_{ar}$) and (a2) three objectives ($\mathcal{J}_{dr},\mathcal{J}_{ar},\mathcal{J}_{dcr}$). (b) The ideal population dynamics in the absence of any error $(\epsilon,\eta,\kappa)=0$ with (b1) the ADR pulse and (b2) the SR pulse.}
\end{figure}

Multiobjective optimization aims to simultaneously optimize multiple conflicting objectives. Here, by adding the cost functions $\mathcal{J}_{ar}, \mathcal{J}_{dr}$ simultaneously we can  
identify the possible pulse that is robust against both errors, which is the amplitude-detuninig robust (ADR) pulse. We consider $\mathcal{J}_{ar}, \mathcal{J}_{dr}$ with individual objectives forming a cost-function set for optimization,
\begin{equation}\label{eq15}
\textcolor{black}{{\mathcal{J}_{adr}}= \{\mathcal{J}_{ar}, \mathcal{J}_{dr} \}},
\end{equation}
with all parameters initialized unchanged. 
\textcolor{black}{During each optimization epoch we have to evaluate both $\mathcal{J}_{ar}$ and $\mathcal{J}_{dr}$ to construct the multiobjective cost function $\mathcal{J}_{adr}$, which is minimized via concurrent optimization with non-dominated sorting.} Unlike the one-objective optimization in Sec. III, through iterative computation, 
 there exists a trade-off between the amplitude and detuning objectives which gives rise to a set of optimal solutions instead of one best solution, forming the Pareto Front (PF) as shown in Fig.\ref{ParetoFront}(a1) \textcolor{black}{\cite{Pareto_front_Kang2024}}. For the PF, enhancing one objective without negatively impacting other objectives is unfeasible. Consequently, no single objective strictly outperforms any other in terms of both optimization targets, indicating that all solutions are non-dominated.

Since all solutions on the PF are in principle optimal, the key challenge lies in selecting the most practically valuable solution for implementing the gate. To achieve this, we utilize the EWM, a rigorous objective weighing technique to identify the optimal solution \textcolor{black}{\cite{EWM_Li2022}}. This method automatically assigns weight coefficients to various objectives by evaluating the dispersion degree of each objective function value, thereby avoiding the subjectivity associated with manual weighting. 
Specifically, we can obtain $n=100$ optimal solutions, equal to the number of initial population size. Each solution $i$ can be represented as a point in Fig.\ref{ParetoFront}(a1) with the coordinate $\left(\mathcal{J}_{dr,i}, \mathcal{J}_{ar,i}\right)$ where $\mathcal{J}_{dr,i}$ and $\mathcal{J}_{ar,i}$ are the computed values of two objective indicators. 
In order to establish a complete ordering of all solutions and select the most optimal one, we need to integrate the dual-objective metrics into a single composite score using the weighted method. The determination of weights is the core of the EWM. Based on scores, the global optimization of solutions can be realized through ranking.

To eliminate the influence of different magnitudes, we begin by standardizing the Pareto solutions as $\left(S_{dr,i}, S_{ar,i}\right)$, with
\begin{equation}   S_{k,i}=\frac{\text{Max}[{\mathcal
{J}_{k}}] - \mathcal{J}_{k,i}}{\text{Max}[{\mathcal
{J}_{k}}]-\text{Min}[{\mathcal
{J}_{k}}]}, k\in(dr,ar)
\end{equation}
where $\text{Max}[{\mathcal
{J}_{k}}]$ and $\text{Min}[{\mathcal
{J}_{k}}]$ are respectively the maximal or minimal value of the indicators. 
In the EWM, the weight of an indicator is intrinsically linked to the dispersion degree of its underlying data, which can be quantified by the information entropy. Especially, when an indicator exhibits a higher data dispersion its entropy value is smaller indicating that it carries more effective information and should be assigned with a relatively larger weight. From Fig.\ref{ParetoFront}(a1) 
the weight of $\mathcal{J}_{ar}$ is larger than that of $\mathcal{J}_{dr}$.
Then the entropy value for indicators $\mathcal{J}_{ar}$ and $\mathcal{J}_{dr}$ is defined as
\begin{equation}   
E_{k}=-\frac{\sum_{i=1}^{n}P_{k,i} \times \ln{P_{k,i}}}{\ln{n}},
\end{equation}
where $P_{k,i}={S_{k,i}}/{\sum_{i=1}^{n}S_{k,i}}$ is the normalized Pareto solution. For all Pareto optimal points, if the distribution of $\left(P_{dr,i}, P_{ar,i}\right)$ is more uniform its entropy value is larger leading to a lower weight value. The weight of two indicators can be calculated by
\begin{equation}   
w_{k}=\frac{1-E_{k}}{\sum_{k}(1-E_{k})}.
\end{equation}

Depending on the EWM, we determine the weight coefficients of two objectives, yielding $w_{ar}=0.8994$ and $w_{dr}=0.1006$. Finally, the single composite score of each optimal solution is obtained via a weighted summation
\begin{equation}   
Score_{i}=w_{ar}\times P_{ar,i}+w_{dr}\times P_{dr,i}\,
\end{equation}
in which the highest score value is selected for designing the ADR pulse with two minimal objective values $\left( \mathcal{J}_{dr},\mathcal{J}_{ar}\right)=\left(3.65\times10^{-4}, 1.29\times10^{-5}\right)$, as marked by a red star in Fig.\ref{ParetoFront}(a1).

We compare the performance of the ADR pulse with other pulses (TO, DR and AR) in Fig.\ref{different_pulse} (a) and (b), as a function of the amplitude $\epsilon$ and detuning $\eta$ errors.
As expected, the ADR pulse performs best which shows a significant robustness against both amplitude and detuning deviations. Note that, the ADR pulse has lowered the infidelity to the level of $\sim 10^{-4}$ or below for the $\epsilon$ error, to $10^{-4}\sim 10^{-3}$ for the $\eta$ error. The population transfer of scheme with the ADR pulse is displayed in Fig.\ref{ParetoFront}(b1). We find the ADR pulse spends a relatively long time in the Rydberg state ($\mathcal{J}_{dcr}={0.5030}$ $\mu$s) and is thus more affected by the decoherence (typically $\sim 10^{-3}$) from its spontaneous decay and dephasing, which allows us to proceed with a three-objective optimization for a super-robust (SR) pulse.

\subsection{SR pulse with decoherence error}

Analogously to the ADR pulse we add the third optimization objective $\mathcal{J}_{dcr}$ into the cost function set, aiming at suppressing the time-spent in the Rydberg state during the cyclic evolution, which is 
\begin{equation}\label{eq11}
\textcolor{black}{\mathcal{J}_{sr}= \{\mathcal{J}_{ar},\  \mathcal{J}_{dr},\ \mathcal{J}_{dcr} \}}
\end{equation}
called the SR pulse, in which $\mathcal{J}_{dcr}$ defined by
\begin{equation}\label{eq10}
\textcolor{black}{\mathcal{J}_{dcr}=\sum_{q=0,1}\int^{\tau}_{0}\vert\langle r \vert \psi_q(t)\rangle|^2dt}
\end{equation}
presents the time spent (in unit of $\mu$s) for the input state in Rydberg state $|r\rangle$. \textcolor{black}{Here, $\psi_q(t) =e^{-i\int_{0}^{t}H(t')dt'}|q\rangle$ is the instantaneous wave function calculated by $t$ time evolution from the initial state $|q\rangle$. Summing over the integrated population $\int^{\tau}_{0}\vert\langle r \vert \psi_q(t)\rangle\vert^{2}dt$
with different input states $q=0,1$ corresponds to the total duration for arbitrarily instantaneous state staying in $\vert r \rangle$.} Remarkably, the robustness to the decoherence error $\kappa$ is sufficiently quantified by minimizing the time integral of population in the Rydberg state when both $\epsilon$ and $\eta$ are ignored.

We further use the cost function set $\mathcal{J}_{sr}$ involving three individual objectives to find the SR pulse. Similar to the two-objective case, the inherent competitive relationship among three objectives remains, arising greater complexity because PF behaves as a three-dimensional surface as demonstrated in Fig.\ref{ParetoFront}(a2). We still employ the EWM as a decision-making strategy to identify the most suitable solution, resulting in the assigned weights for the three objectives {$w_{dr}=0.2666$, $ w_{ar}= 0.4248 $ and $w_{dcr}=0.3086$}. {And the single composite score of each optimal solution turns into}
\begin{equation}
Score_{i}=w_{ar}\times P_{ar,i}+w_{dr}\times P_{dr,i}+w_{dcr}\times P_{dcr,i}.
\end{equation}
The selected point as marked (Fig.\ref{ParetoFront}a2) has a coordinate of {$(\mathcal{J}_{dr}, \mathcal{J}_{ar}, \mathcal{J}_{dcr} )=(0.0012, 0.0007, 0.2782
\mu s)$}. All optimized parameters are summarized in Table \ref{tab:table1}. The ideal population evolution of input state $|0\rangle$ with the SR pulse is displayed in Fig.\ref{ParetoFront}(b2), evidently outperforming the case with ADR pulse with a smaller time spent ($0.2782\mu s<0.5030 \mu s$) in the Rydberg state. A similar behavior is obtained for $|1\rangle$ due to $\Omega_0(t)=\Omega_1(t)$ (not shown).

\section{Super-robust holonomic quantum gates} 

\begin{figure}
\includegraphics[width=3.4in]{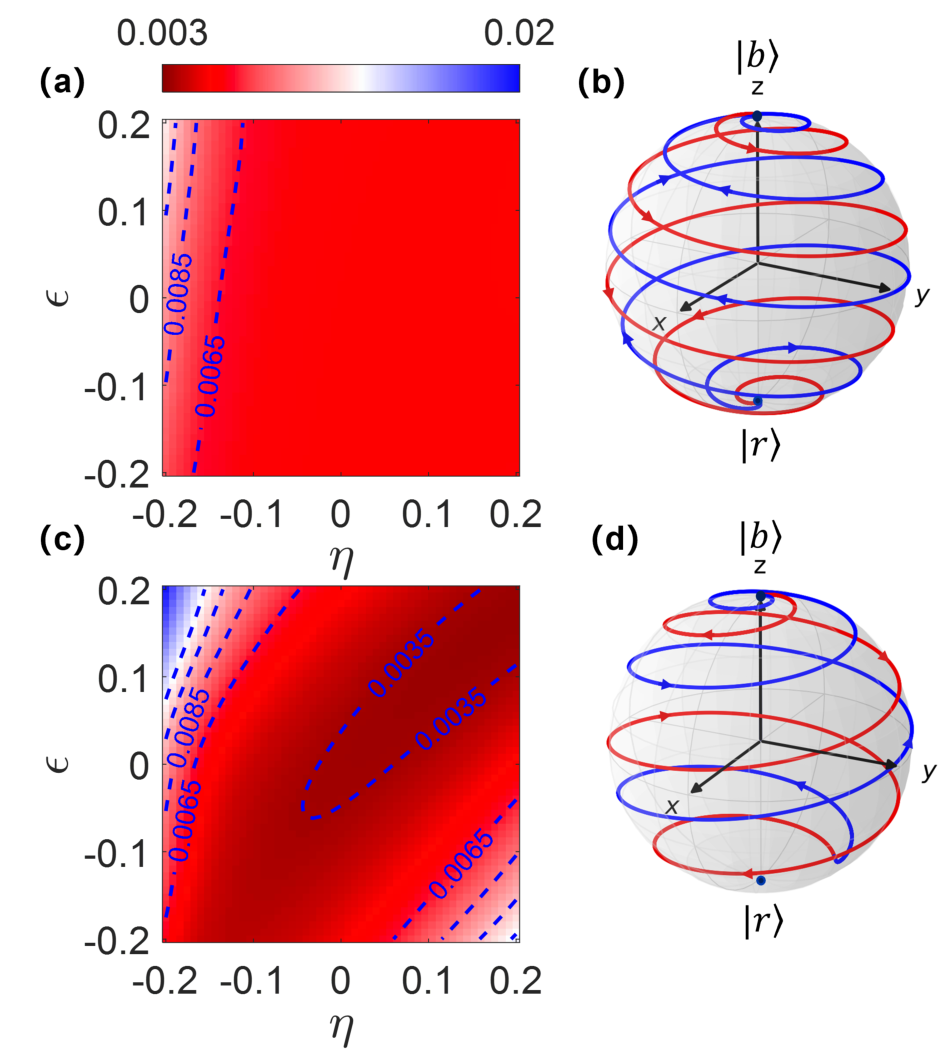}
\caption{\label{SingleQubit} For a practical simulation, the gate infidelity as a function of both $\eta$ and $\epsilon$ in the presence of decoherence $\kappa/2\pi=3.0$ kHz using (a) the ADR pulse and (c) the SR pulse. The corresponding evolution path of $|\mu_2(t)\rangle$ for the $X$ gate on the Bloch sphere are shown in (b) and (d).}
\end{figure}

\subsection{Single-qubit $X$ gate}

We utilize the robust pulses (ADR, SR) under multiobjective optimization, to assess the gate performance in a realistic setup. Now we assume two encoding states 
$\vert 0 \rangle=\vert5S_{1/2}, F=1,m_{F}=0\rangle$ and $\vert 1\rangle=\vert 5S_{1/2},F=2, m_{F}=0\rangle$, which can be driven by a two-photon transition to the Rydberg state $\vert r\rangle=\vert73S_{1/2},m_{J}=-1/2\rangle$ with a finite decoherence rate $\kappa$, considered for $^{87}$Rb atoms \textcolor{black}{\cite{Energy_level_Li2022}}. The results are summarized in Fig.\ref{SingleQubit}. We first consider the gate performance with the ADR pulse, with respect to both $\epsilon$ and $\eta$ errors as shown in (a-b). It is evident that the ADR pulse requires a spiral-like and longer path to implement the cyclic evolution, greatly differing from the TO pulse (Fig.\ref{modone}b). Since the total operation time $\tau$ is fixed it leads to a larger laser amplitude. Moreover, we find the gate with the ADR pulse reveals a better insensitivity to the deviation of $\epsilon$ and $\eta$ errors. The gate infidelity values almost sustain constant within the wide error range of $\eta\in[-0.2,0.2]$ and $\epsilon\in[-0.2,0.2]$, which also arise a relatively poor fidelity \textcolor{black}{$\sim0.9944$} even if $\epsilon=\eta=0$. That can be understood by the spiral-like evolution path of dark state $|\mu_2(t)\rangle$ on the Block sphere which spends a longer time ($\mathcal{J}_{dcr}=0.5030$ $\mu$s) in the lossy Rydberg state $|r\rangle$, leading to a serious decoherence error at the level of \textcolor{black}{$\sim 5.6\times 10^{-3}$}.

When it comes to the SR pulse, we find that it can effectively balance the influence of all three error sources that are competing objectives, by reducing the dominance of the decoherence effect. As a consequence, the impact of control errors ($\epsilon$ and $\eta$) becomes more pronounced, making the gate performance more sensitive to the deviations of them (Fig.\ref{SingleQubit}c). Notably, this sensitivity does not compromise the advantage of the SR pulse. Over a wide range of control errors, the SR pulse outperforms the ADR pulse by having a larger gate fidelity above \textcolor{black}{0.9965}. These results can be attributed to the third objective $\mathcal{J}_{dcr}$ introduced in the multiobjective optimization. As illustrated in Fig. \ref{SingleQubit}d, we note that as compared to the ADR case, the optimized path of the SR pulse can be largely shortened alongside with a big suppression of the time-spent in the Rydberg state $|r\rangle$ (the decoherence error reduces to \textcolor{black}{$3.4\times 10^{-3}$}), which thereby results in an enhanced gate fidelity. Based on the comparison we stress that the SR pulse is the best choice for a practical system in the presence of both control ($\epsilon,\eta$) and decoherence ($\kappa$) errors, presenting stronger robustness in NHQC.

\subsection{Two-qubit CNOT gate}

\begin{figure}
\includegraphics[width=3.4in]{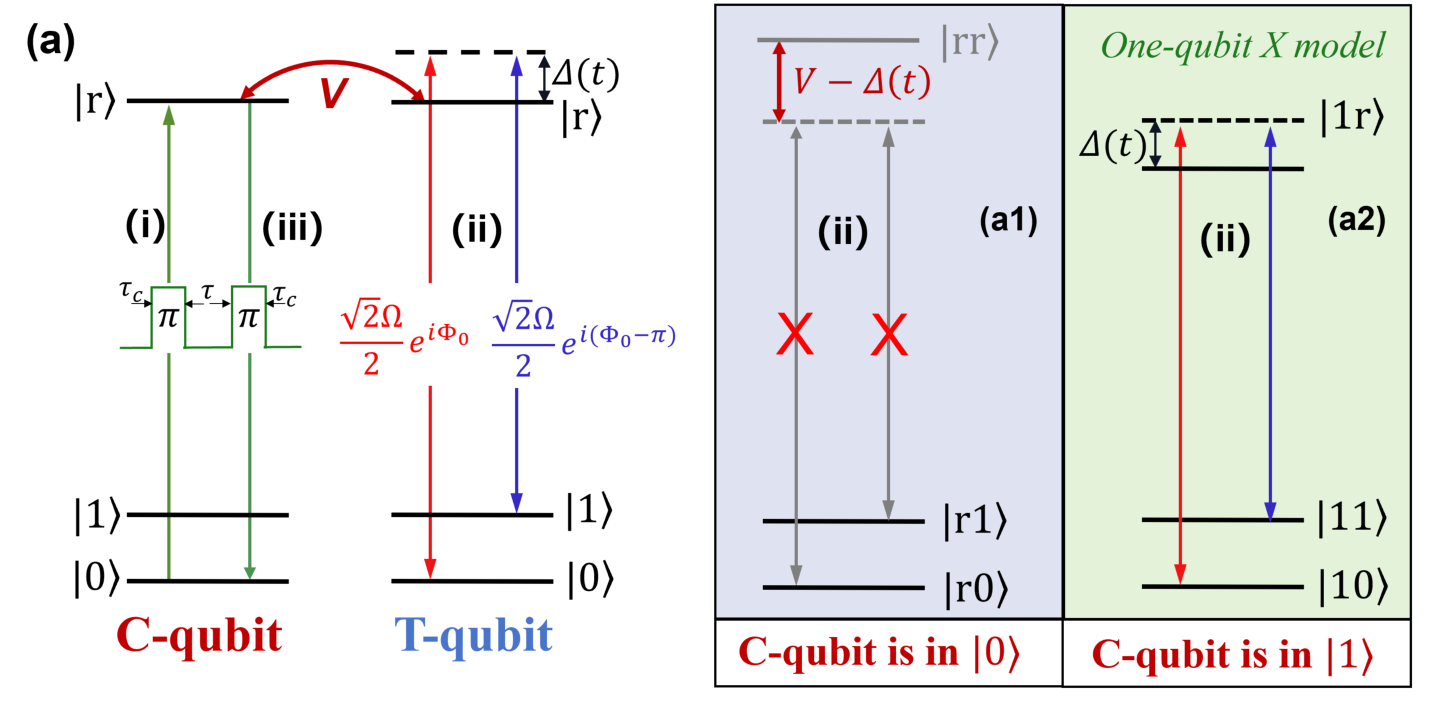}
\caption{\label{modeltwoQubit} Illustration of the two-qubit CNOT gate. (a) Except for the setup of target qubit same as in Fig.\ref{modone}a, 
the control qubit is driven by
$\pi$ pulses in steps (i) and (iii) between $|0\rangle$ and $|r\rangle$. $V$ denotes the Rydberg interaction strength that is of van der Waals type. Insets (a1-2): During step (ii), the target qubit serves as the single-qubit $X$ gate model when the control qubit is in $|1\rangle$, and it remains unexcited due to strong blockade effect $V\gg \Omega$ once the control qubit is in $|0\rangle$.}
\end{figure}

Our robust pulses also enable the implementation of nontrivial two-qubit quantum gates for universal quantum computing. Here we pay attention to realizing the CNOT gates by incorporating with a control qubit (C-qubit) into the $X$ gate model (T-qubit) being one of the typical examples \textcolor{black}{\cite{CNOT_Maller2015}}. To preserve the mechanism of nonadiabatic holonomic quantum gates we adopt a standard $\pi$-gap-$\pi$ approach (Fig.\ref{modeltwoQubit}a). Briefly, when the \textcolor{black}{C-qubit} is initially in state $|0\rangle$ (Fig.\ref{modeltwoQubit}a1) it is excited to the Rydberg level $|r\rangle$ by the first $\pi$ pulse and the induced blockade interaction $V$ acting on $|rr\rangle$ would prevent the $\Omega(t)$ pulses on the \textcolor{black}{T-qubit} from having any effect. Finally the second $\pi$ pulse returns it to the ground state. While, when the \textcolor{black}{C-qubit} is initially in $|1\rangle$ keeping an idler (Fig.\ref{modeltwoQubit}a2), this corresponds to a standard one-qubit $X$ gate model benefiting from the use of robust pulses (see Sec. VA).
The three individual steps in $\pi$-gap-$\pi$ approach can be understood as follows.

\textit{Step} (i). Turn on the $\pi$ pulse $\Omega_c$ on the C-qubit in which the Hamiltonian is
\begin{equation}
H_{c}=\frac{\Omega_{c}}{2}\vert r\rangle\langle 0\vert+H.c.
\end{equation}
with $\Omega_{c}$ being the Rabi frequency that requires $\int_{0}^{\tau_c}\Omega_{c}dt=\pi$. Here we set $\Omega_{c}/2\pi=10$ MHz arising the pulse duration $\tau_c=0.05$ $\mu$s. Note that we ignore the control errors ($\epsilon,\eta$) on the C-qubit because our pulses are not explicitly robust to those errors since the \textcolor{black}{C-qubit} merely serves as the control knob that does not participate in NHQC.

\textit{Step} (ii). Turn off the $\pi$ pulse in step (i) and turn on the modulated robust lasers on T-qubit with the target Hamiltonian $H(t)$ given in Eq. (\ref{eqfull}). Here we introduce the interaction Hamiltonian
\begin{eqnarray}
H_{V}=V\vert rr\rangle\langle rr\vert
\end{eqnarray}
presenting the case when two atoms are both in Rydberg state $|r\rangle$.
$V=C_{6}/R^{6}$ denotes the interaction strength with the dispersion coefficient $C_{6}/2\pi=1416$ GHz$\cdot\mu m^{6}$ \textcolor{black}{\cite{ARC_ibali2017}}, which makes us arrive at $V/2\pi\approx 346$ MHz for a realistic value of the two-atom distance about $R\approx4.0$ $\mu$m. \textcolor{black}{Despite the technical challenges, such strong interaction strength has been demonstrated in recent experiment when the separation $R$ is far smaller than the blockade radius \cite{Nature_Evered2023}.}

\textit{Step} (iii). Turn off the lasers on T-qubit and at the same time turn on the second $\pi$ pulse with a same Rabi frequency $\Omega_{c}$ on the C-qubit, i.e. $\int_{\tau_c+\tau}^{2\tau_c+\tau}\Omega_c dt=\pi$, ensuring the return of the control atom. Then the overall gate time is $2\tau_c+\tau=1.1$ $\mu$s.
Depending on these settings, 
the total Hamiltonian can be described as
\begin{eqnarray}
H_{tot}=H_{c}\otimes I
+I \otimes  H+H_{V}
\end{eqnarray}
with $I=\vert0\rangle\langle0\vert+\vert1
\rangle\langle1\vert$ the identity matrix and $H$ the single-qubit Hamiltonian according to Eq.(\ref{eqfull}). Based on the conditional initialization one could define the whole evolution operator for a two-qubit CNOT gate which is
\begin{eqnarray}
\mathcal{U}=\vert 0\rangle_{c}\langle0\vert\otimes I +\vert 1 \rangle_{c}\langle1\vert \otimes U(\pi/2,\pi,\pi).
\end{eqnarray} 

\begin{figure}
\includegraphics[width=3.4in]{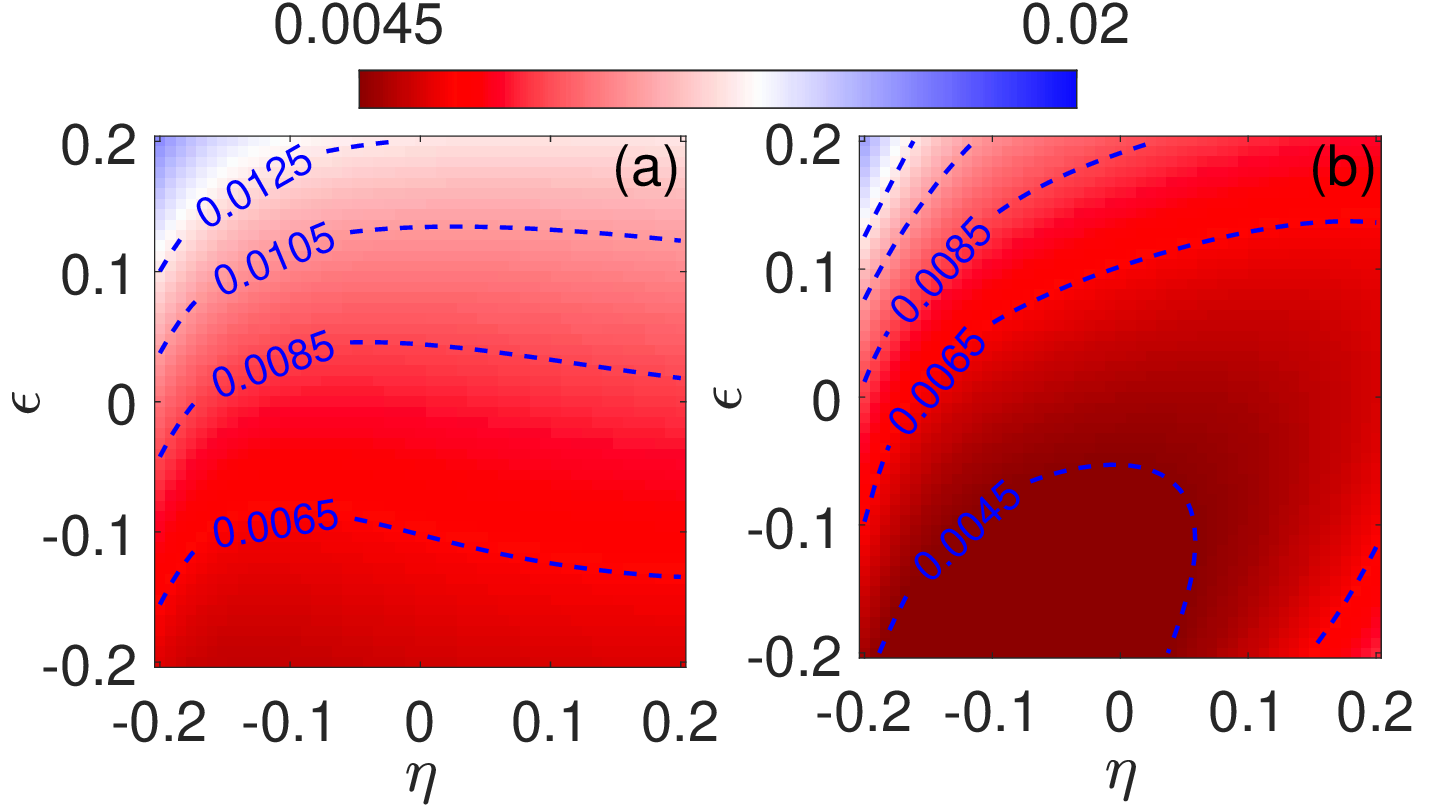}
\caption{\label{TwoQubit} Robustness of two-qubit CNOT gate infidelities against the control errors $\eta$ and $\epsilon$ by utilizing (a) the ADR pulse and (b) the SR pulse for the T-qubit. The decoherence rate is $\kappa/2\pi$ = 3.0 kHz considered for both C- and T-qubits.}
\end{figure}

We test the two-qubit gate performance with two robust pulses (ADR and SR) where the C-qubit only suffers from the decoherence error $\kappa$ serving as a control knob. The results are summarized in Fig.\ref{TwoQubit}. First, we realize that the decoherence effect due to the excitation of the C-qubit during a $2\tau_c$ duration, will induce an extra decoherence error at the level of \textcolor{black}{$\sim 10^{-3}$} resulting in a lower gate fidelity
compared with the single-qubit $X$ gates (Fig.\ref{SingleQubit}). Moreover, the imperfect blockade error that is absent in single-qubit gates \textcolor{black}{\cite{Blockade_error_Shi2017}}, would cause the gate infidelity more sensitive to the control error deviations. Especially, when both $\eta$ and $\epsilon$ are negative which can improve the strength of Rydberg blockade by satisfying $V-(\Delta(t)+\eta)\gg \Omega(t)+\epsilon$, the SR pulse performs best with the gate fidelity more than \textcolor{black}{0.9955}  (Fig.\ref{TwoQubit}b). 
Furthermore, this multiobjective optimization framework can be readily extended to search for more robust pulses, to be used in the implementation of various kinds of two-qubit quantum gates.

\section{Conclusion} 

%Non-adiabatic holonomic quantum gate operation features a perfect gate fidelity in the absence of any error, enabled by its fundamental parallel transport and cyclic evolution conditions. Yet in practice, the inevitable control and environmental-induced decoherence errors in quantum manipulation will decrease the gate fidelity making it very sensitive to the type of errors that occur.

We propose a universal multiobjective optimization protocol for nonadiabatic holonomic quantum gates which achieve simultaneous robustness against three dominant error sources: the amplitude fluctuations, the detuning deviations and the Rydberg decoherence. Different from the traditional one-objective optimization strategies that target at maximizing gate fidelity in the absence of any error \textcolor{black}{\cite{Modulated_Pulses_Fidelity_Chang2023}} or enhancing the robustness to one specific type of the errors \textcolor{black}{\cite{One_objective_Robust_Xu2017}}, our protocol minimizes multiple conflicting objectives simultaneously giving rise to a better trade-off even if these multiple error sources have extremely different magnitudes of impacts. The resulting robust pulses can strongly suppress the errors from amplitude and detuning deviations, and meanwhile reduce the time-spent on the auxiliary Rydberg state giving rise to a smaller decoherence error. Such advantages enable the optimized gates to exhibit superior fidelity and robustness in practice.

Furthermore, the presented optimization framework for one-qubit $X$ gates and two-qubit CNOT gates, is readily generalizable to other types of holonomic quantum gates, even for a multiqubit setting with more objectives \textcolor{black}{\cite{Multiqubit_gate_Khazali2020}}. Finally, we emphasize that our approach enabling a precise quantification of the relative fidelity contributions from distinct error sources,  
could also be adapted to understand the trade-offs in error budget for quantum gates, bringing us one-step closer to identifying which error source can contribute to improve the gate fidelity in practice.

% We present a novel optimization protocol for constructing nonadiabatic holonomic quantum gates with significant robustness. Different from the traditional one-objective optimization that targets at suppressing the average gate error \textcolor{black}{\cite{Modulated_Pulses_Fidelity_Chang2023}} or one type of the errors \textcolor{black}{\cite{One_objective_Robust_Xu2017}}, 
% our multiobjective-based optimization algorithm could minimize multiple conflicting objectives simultaneously giving rise to a better trade-off even if these multiple error sources have extremely different magnitudes of impacts. We observe that our robust pulses can strongly suppress the errors from amplitude and detuning deviations, and meanwhile reduce the time-spent on the auxiliary Rydberg state giving rise to a smaller decoherence error. 
% Note that the ideas introduced here for one-qubit $X$ gates and two-qubit CNOT gates, also generalize to other types of holonomic quantum gates even for a multiqubit setting with more objectives \textcolor{black}{\cite{Multiqubit_gate_Khazali2020}}. Finally, we emphasize that the approach could also be adapted to understand trade-offs in the error budget for quantum gates bringing us closer to identifying which error source can contribute to improve the gate fidelity in practice. 

\begin{acknowledgments}

We acknowledge financial support from the National Natural Science Foundation of China under Grants Nos. 12174106, 11474094 and 11104076), the Natural Science Foundation of Chongqing under Grant No. CSTB2024NSCQ-MSX1117, the Shanghai Science and Technology Innovation Project under Grant No. 24LZ1400600, and the Science and Technology Commission of Shanghai Municipality under Grant No.18ZR1412800.

\end{acknowledgments}

\appendix

\nocite{*}


%\bibliography{apssamp}

\input{apssamp.bbl}% Produces the bibliography via BibTeX.

\end{document}